\documentclass[a4paper,11pt,dvips]{article}
\usepackage{graphicx}
\usepackage{amsmath}
\usepackage{amssymb}
\usepackage{latexsym}
\usepackage[colorlinks]{hyperref}
\usepackage{color}
\usepackage{float}
\usepackage{cite}
\usepackage{makeidx}
\usepackage{colortbl}

\newcommand{\bra}{\begin{array}}
\newcommand{\era}{\end{array}}
\newcommand{\beq}{\begin{equation}}
\newcommand{\eeq}{\end{equation}}
\newcommand{\beqar}{\begin{eqnarray}}
\newcommand{\eeqar}{\end{eqnarray}}

\def\BC{\bb C}
\def\_\BC{\bbi C}

\def\( {\left(}
   \def\) {\right)}
\def\[ {\left[}
\def\] {\right]}
\def\no2 {{\textstyle{n\over 2}}}

\def\dag {{\dagger}}

\newcommand{\om}{\omega}

\newcommand{\lam}{\lambda}
\newcommand{\si}{\sigma}

\newcommand{\be}{\beta}

\newcommand{\ga}{\gamma}

\newcommand{\al}{\alpha}

\newcommand{\da}{\dagger}

\newcommand{\ov}{\over}

\newcommand{\sq}{\sqrt}

\newcommand{\lb}{\label}

\begin{document}
\begin{titlepage}
\setcounter{page}{1}
\renewcommand{\thefootnote}{\fnsymbol{footnote}}

\begin{flushright}
ucd-tpg:1204.03\\
\end{flushright}

\vspace{5mm}
\begin{center}

{\Large \bf {Confined System with Rashba Coupling \\ in Constant Magnetic Field}}

\vspace{5mm}
{\bf Mohammed El Bouziani}$^{a}$,
{\bf Rachid Hou\c{c}a}$^{a}$ and
{\bf Ahmed Jellal$^{a,b,c}$\footnote{\sf ajellal@ictp.it -- a.jellal@ucd.ac.ma}}

\vspace{5mm}

{$^{a}$\em Theoretical Physics Group,  
Faculty of Sciences, Choua\"ib Doukkali University},\\
{\em PO Box 20, 24000 El Jadida,
Morocco}

{$^b$\em Physics Department, College of Science, King Faisal University,\\
PO Box 380, Alahsa 31982, Saudi Arabia}

{$^c$\em Saudi Center for Theoretical Physics, Dhahran, Saudi Arabia}

\vspace{3cm}

\begin{abstract}

We study a  two dimensional system of electrons with Rashba
coupling
in the constant magnetic field $B$
and confining potential.
We algebraically diagonalize the corresponding Hamiltonian
to end up with the solutions
of  the energy spectrum. In terms of two kinds of operator we
construct two symmetries and discuss the filling of the shells with electrons
for strong and weak $B$. Subsequently,
 we show that our system is sharing
 some common
 features with quantum optics
 where the exact operator solutions for the
 basics Jaynes-Cummings variables are derived from our results. An interesting limit
 is studied and the corresponding quantum dynamics is recovered.

 \vspace{3cm}



\end{abstract}
\end{center}
\end{titlepage}


\section{Introduction}

The spin-orbit coupling, which couples the electron spin and its orbital motion,
 has been the subject of several theoretical and experimental research~\cite{zutic}.
This opens the door to developing a new generation of electronic spin (spintronics)
and presents a new branch of physics of semiconductors.
  It puts the spin of the electron at the center of interest
  and exploits the spin-dependent
electronic properties of magnetic materials and semiconductors.
  The underlying basis for this new electronics is the intimate connection between the electron charge and spin.
  A crucial implication of this relationship is that one can have access to spin through the spin property of the electron orbital in the solid.
  The link between the electron charge and spin is expressed by the spin-orbit interaction in semiconductors.

Novel spin properties arise from the interplay
between Rashba spin splitting \cite{rashba} and further confinement
of two dimensional (2D) electrons in quantum wires \cite{mireles,hausler,governale1,schaupers},
rings \cite{splett,foldi} or dots \cite{vosko,governale2,valin1,bulgakov1,valin2,destefani,zaitsev}. Spin-orbit coupling
has also been shown to affect the statistics of energy levels
and eigenfunctions as well as current distributions \cite{bulgakov2,berggren}.
The interplay between spin-orbit coupling and external
magnetic fields was analyzed theoretically using random
matrix theory \cite{aleiner}.
With this respect,
Schliemann~\cite{schliemann} studied the cyclotron motion and magnetic focusing in
semiconductor quantum wells with spin-orbit coupling. More precisely,
 the ballistic motion of electrons in III-V semiconductor quantum wells with Rashba spin-orbit coupling
in a perpendicular magnetic field was investigated. Taking into account the full quantum dynamics of the problem,  the modifications
of classical cyclotron orbits due to spin-orbit interaction was explored
and the analogy to Jaynes-Cummings was established. {Also we mention some works
that have contributed to the understanding of magnetic effects and spin-orbit coupling in quantum wires \cite{group}.}

On the other hand, Jaynes-Cummings  model is describing
the basic interaction of a two-level atom and
quantized  field, which  is also the cornerstone for the
treatment of the interaction between light and matter in quantum
optics~\cite{jaynes}. It can be used to explain many quantum phenomena,
such as the collapses and revivals of the atomic population
inversions, squeezing of the quantized field and the atom-cavity
entanglement. Recent experiments showed that Jaynes-Cummings model
can be implicated in quantum-state engineering and quantum
information processing, e.g. generation of Fock states~\cite{bertet}
and entangled states~\cite{osnaghi}, and the implementations of quantum
logic gates~\cite{monroe}, etc. Originally, Jaynes-Cummings model is physically
implemented with a cavity quantum electrodynamics system,
see for instance~\cite{rempe}.
Certainly, there has been also interest to
realize Jaynes-Cummings model with other physical
systems. A typical system is a cold ion trapped in a Paul trap and
driven by classical laser beams~\cite{blockey,leibfried} where the
interaction between two selected internal electronic levels and
the external vibrational mode of the ion can be induced.

Motivated by different investigations  cited above
and in particular~\cite{schliemann,gazeau}, we develop our proposal to deal with different
issues. For this, we consider a  2D system in the presence of an external magnetic field $B$
and study the quantum dynamics. But, we include the parabolic potential to confine our system and
 Rashba interaction to make contact with quantum optics.
Through the Weyl-Heisenberg symmetries, we
obtain
the solutions of the energy spectrum
and construct the algebra $su(2)$ as well as $su(1,1)$.
By considering strong and week $B$, we show that our system
reduces to the Landau problem for the first case.
By using
the Heisenberg picture,  we
derive
two copies of the Jyanes-Cummings model oscillating with different
frequencies.
Finally, we recover the results of without confinement case~\cite{schliemann}  in simply way
to conclude that our findings are general and deserve different
extensions.

The present paper is organized as follows. In section 2 we formulate our problem
by setting the Hamiltonian and choosing the convenient gauge.
In section 3, we introduce a series of annihilation and creation operators
to diagonalize our Hamiltonian, which serves to determine explicitly the exact eigenvalues
and eigenstates. We construct two symmetries and analyze the system behavior by distinguishing
the strong and {weak} magnetic field cases in section 4. We establish a link with Jaynes-Cummings
model and make different comments  in section 5. Moreover,  to show the relevance of our results
we study a liming case.  Finally, we close by
concluding our work and giving some perspective.

\section{Solutions of energy spectrum}

We start by formulating our problem to end up
with the appropriate Hamiltonian describing the
system under consideration. Subsequently, we use
the algebraic approach to determine explicitly
the eigenvalues as well as the eigenstates. These
will be used to discuss the possibility to fill
the shells with electrons when the magnetic field
is strong and weak.

\subsection{Hamiltonian formalism}
We consider a system of electrons
in the presence of a constant magnetic field $\vec B= B \vec e_z $
and  confining potential.
By taking into account of the Rashba spin-orbit coupling
and Zeeman effect, the Hamiltonian for
a single electron reads as
\beq\lb{011}
H={\vec\pi^2\over 2m}+{1\over 2} m\om_{0}^2\left(x^2+y^2\right)+\lambda
\left(\pi_x\si_y-\pi_y\si_x\right)+{1\over 2}g\mu_BB\si_z
\eeq
where $\vec{\pi}=\vec{p}+{e\over c}\vec{A}$ is the conjugate
momentum
and $\vec{A}$ is the vector potential.
$\lambda$ is the Rashba coupling parameter,  $g$ is the Land\'e-factor,
$\mu_B$ is the Bohr magneton and $\vec{\si}=\left(\si_x,\si_y,\si_z\right)$ are the  Pauli matrices.
To proceed further, we choose the symmetric Landau gauge
\beq
\vec{A}={B\over2}\left(-y,x,0\right)
\eeq
and write the Hamiltonian
(\ref{011}) as
\beqar\lb{HH}
H &=&{1\over 2m}\left[\left(p_x-{eB\over 2c}y\right)^2+\left(p_y+{eB\over 2c}x\right)^2\right]+{1\over 2}
m\om_{0}^2(x^2+y^2)\nonumber\\
&& +
\lambda\left[\si_y\left(p_x-{eB\over 2c}y\right)-\si_x\left(p_y+{eB\over 2c}x\right)\right]+{1\over
2}g\mu_BB\si_z.
\eeqar

The algebraic structure of the above Hamiltonian is easily displayed if
we adopt the method of separation of Cartesian variables. This process suggests  to decompose
\eqref{HH} into four parts
\beq\lb{222}
H= H_{F} + H_{R} + \frac{1}{2}\om_c L_z + {1\over 2}g\mu_BB\si_z
\eeq
such that the free part takes the form
\beq
H_{F} = \left({p_x^2\over2m}+{1\over8}m\om^2x^2\right)+\left({p_y^2\over2m}+{1\over8}m\om^2y^2\right)
\eeq
and the Rashba coupling in magnetic field is given by
\beq\lb{ras}
H_{ R} =\lambda\left[\si_y\left(p_x-{eB\over 2c}y\right)-\si_x\left(p_y+{eB\over 2c}x\right)\right].
\eeq
where $\om_c={eB\over mc}$ is the cyclotron
frequency,
$L_z= xp_y-yp_x$ is the angular momentum and we have
set the new frequency as $\om=\sq{\om_c^2+4\om_0^2}$. To close this part, we emphasis that (\ref{222}) 
{splits} into two independent harmonic
oscillator Hamiltonian's supplemented by  the angular momentum and Rashba spin-orbit coupling added
to Zeeman term. This convenient form of the Hamiltonian will help us to end up with its diagonalization
in the simple way.

\subsection{Solution through Weyl-Heisenberg symmetries }

We introduce the standard machinery and techniques to get the solutions of energy spectrum
of the Hamiltonian (\ref{222}).
Instead of directly using the oscillator annihilation
operators
\beq
a_x={1\over \sqrt{2}}\left({x\over l_0}+{il_0\over
\hbar}p_x\right),\qquad a_y={1\over \sqrt{2}}\left({y\over
l_0}+{il_0\over \hbar}p_y\right)
\eeq
we work with two new ones,
which are linear superposition of $a_x$ and $a_y$, such as
 \beq
a_d={1\over \sq{2}}\left(a_x-ia_y\right), \qquad a_g={1\over
\sq{2}}\left(a_x+ia_y\right)
\eeq
where $l_0=\sq{2\hbar\over
m\om}$ being the magnetic length. Note that,  $a_d$ and $a_g$ are
bosonic operators and satisfy the relation commutations
\beq
[a_d,a^{\dagger}_d]=1=[a_g,a^{\dagger}_g]
\eeq
and other relations vanish.
{From} the above operators, one can
obtain the useful identities for the conjugate momentum
 \beq\lb{pieq}
 \pi_x = {\hbar\over
2il_0}\left[l_1\left(a_d-a_d^{\dagger}\right)+l_2\left(a_g-a_g^{\dagger}\right)\right],
 \qquad  \pi_y={\hbar\over 2l_0}\left[l_1\left(a_d+a_d^{\dagger}\right)-l_2\left(a_g+a_g^{\dagger}\right)\right]
 \eeq
 as well as for the positions
 \beq
 x = \frac{l_0}{2}\left(a_d+a_d^\dag + a_g+ a_g^\dag\right), \qquad
 y= \frac{l_0}{2i}\left(-a_d+a_d^\dag + a_g - a_g^\dag\right)\lb{xyrep}
\eeq
where we have set $l_1=\left(1+{l_0^2\over 2l^2}\right)$,  $l_2=\left(1-{l_0^2\over
  2l^2}\right)$ and $l^2={\hbar\over m\om_c}$. 
These algebraic structures will play a crucial role in solving different issues
and more precisely in diagonalizing different Hamiltonian's entering in
the game.

We start by writing the Rashba Hamiltonian (\ref{ras}) in terms of the
annihilation and creation operators introduced above. Indeed, we have
\beq\lb{h_r}
 H_R=H_{R}^g+H_{R}^d
\eeq
where these two parts are given by
\begin{equation}
 H_R^g=\lambda\sq{{m\hbar\om\over2}}
\\ \left(1-{\om_c\over\om}\right)\left(%
\begin{array}{cc}
  0 & a_g\\
 a_g^{\dagger} & 0 \\
\end{array}%
\right),\qquad  H_R^d=-\lambda\sq{{m\hbar\om\over2}}
\\ \left(1+{\om_c\over\om}\right)\left(%
\begin{array}{cc}
  0 & a_d\\
 a_d^{\dagger} & 0 \\
\end{array}%
\right).
\end{equation}
In the same way, we can diagonalize the free Hamiltonian and angular momentum to finally
end up with a new form of the Hamiltonian (\ref{222}). This is
\begin{equation}\lb{H}
H={\hbar\om\over2}\left(a_d^{\dagger}a_d+a_g^{\dagger}a_g+1\right)+{\hbar\om_c\over2}\left(a_d^{\dagger}a_d-a_g^{\dagger}a_g\right)+
H_{R}^g+H_{R}^d+{1\over2}g\mu_BB\si_z.
\end{equation}

To determine the solutions of energy spectrum of the above problem,
we solve the eigenvalue equation
\beq
H \left(%
\begin{array}{cc}
  \psi_1\\
 \psi_2 \\
\end{array}%
\right) = E \left(%
\begin{array}{cc}
  \psi_1\\
 \psi_2 \\
\end{array}%
\right)
\eeq
which gives the eigenvalues
{\beq
E_{n_dn_g} =  \hbar\omega^{+}n_d+ \hbar\omega^{-}n_g \pm \frac{1}{2}\hbar \om^- \sq{8{m\lam^2\ov\hbar\omega}
n_g+1}
 \pm
\frac{1}{2}\hbar \om^+\sq{8{m\lam^2\ov\hbar\omega} n_d+\left(1+\frac{g\mu_BB}{\hbar \om^+}\right)^2}
\eeq}
 where the new frequencies are $\omega^{\pm}=\frac{1}{2}\left(\omega\pm\omega_c\right)$. The corresponding
eigenstates read as
\beq\lb{est}
|n_g,n_d,\si\rangle=u_n^{\pm}|n_g,n_d,\uparrow\rangle+v_n^{\pm}|n_g-1,n_d-1,\downarrow\rangle
\eeq
and we show that the amplitudes parameterizing these states are given by
\beq u_n^{\pm}=\frac{1}{\sqrt{2}}\left(1 \pm
\frac{\hbar\om+g\mu_BB}
{\hbar \om^-\sq{8{m\lam^2\ov\hbar\omega} n_g+1}
+\hbar \om^+\sq{8{m\lam^2\ov\hbar\omega} n_d+\left(1+\frac{g\mu_BB}{\hbar \om^+}\right)^2}
}\right)^{1\ov2}
\eeq
\beq v_n^{\pm}=\frac{\pm i}{\sqrt{2}}\left(1\mp
\frac{\hbar\om+g\mu_BB}
{\hbar \om^-\sq{8{m\lam^2\ov\hbar\omega} n_g+1}
+\hbar \om^+\sq{8{m\lam^2\ov\hbar\omega} n_d+\left(1+\frac{g\mu_BB}{\hbar \om^+}\right)^2}
}\right)^{1\ov2}.
\eeq

Having obtained the solutions of the energy spectrum, let us briefly
discuss how to recover an interesting case from what we generated so
far.  Indeed,
in studying cyclotron motion and magnetic focusing in semiconductor quantum wells with spin-orbit coupling,
Schliemann~\cite{schliemann} introduced the Hamiltonian type \eqref{011} without the confining potential.
Therefore, to recover the corresponding solutions of energy spectrum, we consider the case $ \om=\om_c$
or $\om_0=0$
 in the previous equations. This gives
the eigenvalues
\beq
\varepsilon_{n_d}
=\hbar\om_cn_d\pm\sq{2m\lam^2
\hbar\om_c n_d+{1\ov4}\left(\hbar\om_c+g\mu_BB\right)^2}
\eeq
as well as the eigenstates
 \beq\lb{n}
|n_g,n_d,\si\rangle=u_n^{\pm}|n_g,n_d,\uparrow\rangle+v_n^{\pm}|n_g-1,n_d-1,\downarrow\rangle
\eeq
where the amplitudes are given by
\beqar
u_n^{\pm} &=&\left({1\over2} \pm
{{1\over4}\left(\hbar\om_c+g\mu_BB\right)\over
\sq{2m\lam^2\hbar\om_c n_d+{1\ov4}\left(\hbar\om_c+g\mu_BB\right)^2}}\right)^{1\ov2}\\
v_n^{\pm} &=& (\pm i)\left({1\over2} \mp
{{1\over4}\left(\hbar\om_c+g\mu_BB\right)\over
\sq{2m\lam\hbar\om_{c} n_d+{1\ov4}\left(\hbar\om_{c}+g\mu_BB\right)^2}}\right)^{1\ov2}.
\eeqar
We mention that Schliemann~\cite{schliemann} used the notation $\al=\hbar \lam$ and the quantum number $n=n_d$.
These show clearly that our findings are general and make difference with what obtained
in~\cite{schliemann}. Certainly, this  will play a crucial role
in the forthcoming analysis.

\subsection{Symmetries and shells}

To make comparison with interesting work~\cite{gazeau}
dealing with confined 2D system in magnetic field,
let us introduce two symmetries. These concern the
dynamical symmetries $su(2)$ and $su(1,1)$, which can be realized
in terms of the shell operators. We start with
$su(2)$ where the corresponding generators can be realized
as
\beq
S_+ = a_d^{\dagger} a_g, \qquad S_- = a_g^{\dagger} a_d, \qquad S_z = 
\frac{L_z}{2\hbar}
\eeq
which verify
the commutation relations
\beq
\lbrack S_+,S_- \rbrack = 2S_z, \qquad \lbrack S_z,S_{\pm} \rbrack = \pm S_{\pm}.
\eeq
These give invariant Casimir operator
\beq
{\cal C} = \frac{1}{2} \left(S_+ S_- + S_- S_+\right) + S_z^2 =
\frac{H_F^2}{\hbar^2\om^2} -\frac{1}{4}.
\eeq
Therefore, to a  fixed value $\mu = (n_d+n_g)/2$ of the operator 
$\frac{{H}_F}{\hbar \om} - \frac{1}{2}$ there
corresponds the $(2\mu + 1)$-dimensional unitary irreducible representation (UIR) of $su(2)$ in which the operator $ S_z = 
\frac{L_z}{2\hbar}$ assumes its spectral values in the range $ -\mu \leq \ga = (n_d - n_g)/2
\leq \mu$.

As far as the second symmetry $su(1,1)$ is concerned, we consider the generators
\beq
T_+ = a_d^{\dagger} a_g^{\dagger}, \qquad T_- = a_g a_d, \qquad T_0 =
\frac{{H}_F}{\hbar \om}
\eeq
satisfying the relations
\beq
\lbrack T_+,T_- \rbrack = -2T_0, \qquad \lbrack T_0,T_{\pm} \rbrack = \pm T_{\pm}.
\eeq
The Casimir operator then is given by
\beq
{\cal D} = \frac{1}{2}(T_+ T_- + T_- T_+) - T_0^2
= -\frac{1}{4} \left(\frac{L_z^2}{\hbar^2} -1\right).
\eeq
Similarly to the previous case, to a fixed value $\eta = (n_d-n_g)/2 +1/2 \geq 1/2 $, with
$n_d - n_g = \al \geq 0$,  of the operator
$\frac{1}{2}\left(\frac{L_z}{\hbar}+1\right)$
there corresponds a  UIR of $su(1,1)$ in the discrete
series,  in which the operator
$T_0=\frac{1}{2}\left(\frac{L_z}{\hbar}+1\right) + N_g$
assumes its spectral values in the infinite
range $ \eta, \eta +1, \eta + 2, \cdots$.
Alternatively, to a fixed value $\varrho = -(n_d-n_g)/2 +1/2 \geq 1/2$, with $n_g - n_d = -\al \geq 0$ of the operator
$\frac{1}{2}\left(-\frac{L_z}{\hbar}+1\right)$
there corresponds a  UIR of $su(1,1)$ in the discrete
series,  in which the operator
$T_0=\frac{1}{2}\left(-\frac{L_z}{\hbar}+1\right) + N_d$
assumes its spectral values in the infinite
range $ \varrho, \varrho +1, \varrho + 2, \cdots$.

Below we will see the importance of the both symmetries
introduced above when
we analyze two interesting cases. With these we underline the system
behavior with respect to different limit of the magnetic field $B$.
We start our analysis by considering the
case where $B$ is strong, which is equivalent to
the limit
$\omega_c\gg\omega_0$, and gives the frequencies $\om^+\simeq\om_c$ and $\om^-\simeq 0$.
These  tell us that
the total energy can be
approximated by
\beq
E_{n_d}\simeq\hbar\omega_c \left(n_d\pm{1\ov2}\right) \pm \frac{1}{2}g\mu_BB
 \eeq
where the quantum number is $n_d = 0, 1\cdots$. Now by scaling the energy as
\beq
E_{n_d} \mp \frac{1}{2}g\mu_BB \simeq\hbar\omega_c \left(n_d\pm{1\ov2}\right) =\varepsilon_{n_d}
\eeq
one can realize immediately that our system behaves like a harmonic oscillator
in 2D (Landau problem) and for a given $n_d$  there is an infinite degeneracy
of the Landau levels. We notice that there are {two types} of quantum numbers $n_d+\frac{1}{2}$  and
$n_d -\frac{1}{2}$, which means that we have two {independent sectors of the Hilbert space}. However,
both {sectors} are connected via a linear transformation $n=n_d+1$
that allows to move from one {sector} to another and vice versa.
From symmetry point view, this behavior can be regarded
as our system has ladder states for
a discrete series representations
of the algebra $su(1,1)$ labeled by
$\frac{1}{2}(-\al \pm 1)$ where
$\al=n_d-n_g\leq n_d$ for $\al\leq 0$.

For weak magnetic field, which corresponds to
the limit $\omega_c \ll\omega_0$,
we can approximate the frequencies by
$\om^+=\om^-\simeq\om_0$
and therefore write the total energy as
\beq\lb{appe}
E_{n_g,n_d}\simeq\hbar\omega_0\left(n_d+n_g\pm 1\right) \pm{1\ov2}g\mu_BB.
 \eeq
 To interpret this result let us rearrange it as follows
 \beq
  E_{n_g,n_d} \mp {1\ov2}g\mu_BB \simeq \hbar\omega_0\left(2\lam\pm 1\right) = \varepsilon_\lam
 \eeq
 which shows clearly that our system becomes now invariant under
 the algebra $su(2)$ and therefore each Landau level has a degeneracy
 of order $(2\lam \pm 1)$. Note that, here also we have two UIR
 of dimensions $(2\lam+1)$ and $(2\lam -1)$ for the same algebra
 where the transition between them can be obtained by defining
 $\rho=\lam-1$. Furthermore, from \eqref{appe} we see that we need
 $2(\lam_0 \pm 1) (2\lam_0 \pm 1)$ to fill the shells up to the value $\lam_0$.

 \section{Link with Jaynes-Cummings model}

 Very recently, it appeared beautiful connections between
 different areas of physics.  Among them, we cite the extraordinary
 bridge between condensed matter physics
and high energy physics through the
link of graphene with quantum electromagnetic~\cite{novoselov}.
Also, another connection between massless Dirac electrons
and quantum optics has been established~\cite{jellal}.
These links push and motivate
to look for bridges and establish  contacts
between different systems. For this purpose,
we make contact with another area of physics
by showing how our system can be linked to
quantum optics through a mapping between
the corresponding Hamiltonian and  {Jaynes-Commings} model. This may
help to strength a good knowledge of
different aspects of quantum optics.

\subsection{Equivalence between models}

 To show the relevance of the results obtained so far, we study
the presence and absence of the confining potential cases
 in the Hamiltonian system. For first one, we will
 show that
 our Hamiltonian (\ref{HH}) is formally equivalent to two copies of the
Jaynes-Cummings models for atomic transition in a radiation
field, but oscillating with different frequencies. To do our job, we adopt the same method
used by Ackerhalt and Rzazewski \cite{ackerhalt} in analyzing the operator perturbation theory in the
Heisenberg picture. For second one, we derive the corresponding results
in simple way from our findings.

To proceed further, we need to rearrange our Hamiltonian in order
to deal with each part separately and establish the associate link. For this,
 we start by splitting  (\ref{H})
into two parts
\beq\lb{hams}
H=H_g+H_d
\eeq
where the first one is
\beq\lb{hamg}
H_g={\hbar\om^-}a_g^{\dagger}a_g+ \frac{\hbar\om}{4} +
H_{R}^g+ {1\over4}g\mu_BB\si_z
\eeq
and  the second reads as
\begin{equation}\lb{hamd}
H_d={\hbar\om^+} a_d^{\dagger}a_d +{\hbar\om\over4}+
H_{R}^d+{1\over4}g\mu_BB\si_z.
\end{equation}
Clearly, these two {parts} are completely different because they  are involving different frequencies
and therefore different oscillations.

Let us consider the Hamiltonian (\ref{hamg}) and make contact with  Jaynes-Cummings model. In doing so,
we show that (\ref{hamg}) can be written as
\beq\lb{n}
H_g=\hbar\omega^{-}
M_g-\gamma^-\sigma_z+\zeta(\omega)\left(a_g^\dag\sigma^{-}+a_g\sigma^{+}\right)
\eeq
where the two operators $M_g$ and $\sigma^\pm$  are give by
\beqar
M_g &=& N_g+\sigma^{+}\sigma^{-} +  \frac{\om_c}{2(\om-\om_c)}\mathbb{I}\\
\sigma^{\pm} &=& {1\ov2}\left(\sigma_x\pm i\sigma_y\right)
\eeqar
and we have set the constants
 $\gamma^-$ and $\zeta(\omega)$ as
\beqar
\gamma^- &=& \frac{1}{4} \left(2\hbar\om^- - g\mu_B B\right)\\
\zeta(\omega) &=& \lambda\sq{{m\hbar\om\over2}}
\left(1-{\om_c\over\om}\right).
\eeqar
For later use, it convenient to define an operator as
\beq\lb{cg}
C_g=-\gamma^-\sigma_z+\zeta(\omega) \left(a_g^\dag\sigma^{-}+a_g\sigma^{+}\right)
\eeq
which verifies the commutation relation $[M_g,C_g]=0$. It tells us
that
there are two constants of motion corresponding to the Hamiltonian
(\ref{n}). This will help in studying different dynamics of
the involved operators.

To study the dynamics related to the Hamiltonian \eqref{n}, we introduce
the
Heisenberg equation of motion for the operators $a_g^\dag$ and
$\sigma^{+}$.
These are
\beqar
{d\ov dt}a_{g}^{\dag} &=&{i\ov\hbar}\left[H_g,a_g^{\dag}\right]
\\ 
{d\ov dt}\sigma^{+}&=&{i\ov\hbar} \left[H_g,\sigma^{+}\right].
\eeqar
A
straightforward calculation leads
\beqar
&&\left(i\hbar{d\ov
dt}+\hbar\omega^{-}\right)a_{g}^{\dag}=-\zeta(\omega)\sigma^{+}\lb{444}
\\
&&\left(i\hbar{d\ov
dt}+\hbar \omega^-- 2{\gamma^-}\right)\sigma^{+}= {\zeta(\omega)}a_g^{\dag}\si_z \lb{0}.
\eeqar
Using the relations
$\sigma^+\sigma^+=\sigma^-\sigma^-=0,
\sigma^{+}\sigma_{z}=- \sigma^{+}$ and
$\sigma^{-}\sigma_{z}= \sigma^{-}$,
we show that  (\ref{0}) can be written in terms of the
constant of motion $C_g$ as
\beq\lb{00}
\left(i\hbar{d\ov
dt}+\hbar \omega^-+2C_g\right)\sigma^{+}=\zeta(\omega) a_g^{\dag}.
\eeq
From  (\ref{444}) and (\ref{00}), we derive
the same second order
differential equation for $\sigma^{+}(t)$ and $a_g^{\dag}(t)$. This is given by
\begin{equation}\lb{dd}
\left(i\hbar{d\ov dt}+\hbar\omega^{-}\right)
\left(i\hbar{d\ov
dt}+\hbar \omega^-+2C_g\right)
\left(%
\begin{array}{c}
 \sigma^{+} \\
  a_g^{\dagger} \\
\end{array}%
 \right)= -\zeta^{2}(\omega)\left(%
\begin{array}{c}
  \sigma^{+} \\
  a_g^{\dagger} \\
\end{array}%
\right).
\end{equation}

We are looking for  the quantum dynamics, then we need to solve
(\ref{444}) and (\ref{00}). One way to do is to  find solutions of the form
\beqar
\sigma^{+}(t) &=& e^{i\beta_g^{+}t/\hbar} s_g^{+}+e^{i\beta_g^{-}t/\hbar}s_g^{-} \lb{v}\\
a_g^{\dag}(t) &=& e^{i\beta_g^{+}t/\hbar}l_g^{+}+e^{i\beta_g^{-}t/\hbar} l_g^{-}\lb{vv}
\eeqar
where
$\beta^{\pm}_g$, $l_g^{\pm}$, and $s_g^{\pm}$ are initial time operators.
From the shape of the decomposition  (\ref{444}) and (\ref{00}), one can remark
that they have an analogy with the solution of ordinary harmonic oscillator.
Therefore, the solution will be of the form
$e^{i\beta_g t/\hbar}$ and thus  after substitution  into
(\ref{dd}) gives a second order equation for $\be_g$. This is
\beq
\beta_g^{2}- 2\left(\hbar\omega^{-}+ C_g\right)\beta_g
+\hbar\omega^{-}\left(\hbar\omega^-+2C_g\right) + \zeta^2(\omega)=0.
\eeq
By requiring the condition $[\beta_g,C_g]=0$, we show that
the corresponding solutions under the decomposition forms
\beq
\beta_g^{\pm}=\hbar \om^-+\al_g^{\pm}
\eeq
where $\al_g^{\pm}$ are given by
\beq
\al_g^{\pm}= \hbar \om^-+ C_g\pm\sqrt{C_g^{2}-\zeta^2(\omega)}.
\eeq
 From the decomposition of $a_g^\dag(t)$ and $\si^+(t)$, one can notice that
the operators constants $s_g^{\pm}$ and $l_g^{\pm}$ can be obtained
by fixing $t=0$ in
(\ref{v}) and (\ref{vv}). These give the relations
\beqar
\sigma^{+}(0) &=&s_g^{+}+s_g^{-} \lb{2}\\
a_g^{\dag} (0)&= & l_g^{+}+l_g^{-}. \lb{22}
\eeqar
Injecting the forms (\ref{v})-(\ref{vv}) into (\ref{444}) and
(\ref{00}) to end up with
\beqar
\al_g^{+}l_g^{+}+\al_g^{-}l_g^{-} &=& \zeta(\omega)\left(s_g^{+}+s_g^{-}\right) \lb{1}\\
\al_g^{-}s_g^{+}+\al_g^{+}s_g^{-} &=& \zeta(\omega)\left(l_g^{+}+l_g^{-}\right).
\lb{11}
\eeqar
These can be solved to obtain the initial operators as
\beqar
s_g^{\pm} &=& \frac{\pm\zeta(\omega)
a_g^{\dag}(0)\mp\al_g^{\pm}\si^+(0)}{\al_g^--\al_g^+} \\
l_g^{\pm} &=&  \frac{\mp\zeta(\omega) \si^{+}(0) \pm \al_g^{\mp}a_g^{\dag}(0)}{\al_g^--\al_g^+}.
\eeqar
Combining all to end up with the final solutions of
(\ref{444}) and (\ref{00}). These are
\begin{eqnarray}
\sigma^{+}(t) &=&{e^{i\omega^-t}\ov
\al_g^--\al_g^+}\left\{\left(\zeta(\omega)
a_g^{\dag}(0)-\al_g^+\si^+(0)\right)e^{i{\al_g^{+}t/\hbar}}+\left(-\zeta(\omega)
a_g^{\dag}(0)+\al_g^+\si^+(0)\right)e^{i{\al_g^{-}t/\hbar}}\right\}\lb{3}\\
a_g^{\dag}(t) &=& {e^{i\omega^-t}\ov
\al_g^--\al_g^+}\left\{\left(-\zeta(\omega)
\si^{+}(0)+\al_g^-a_g^{\dag}(0)\right)e^{i{\al_g^{+}t/\hbar}}+\left(\zeta(\omega)
\si^{+}(0)-\al_g^+a_g^{\dag}(0)\right)e^{i{\al_g^{-}t/\hbar}}\right\}.\lb{33}
\end{eqnarray}
They constitute
the exact operator
solutions for the basic Jaynes-Cummings variables, which have
been obtained in~\cite{ackerhalt}.

As far as the second part  (\ref{hams}) is concerned,
we apply the same machinery as before to derive similar results. Indeed,
by introducing the operator
\beq
M_d = N_d+\sigma^{+}\sigma^{-} -  \frac{\om_c}{2(\om+\om_c)}\mathbb{I}
\eeq
and the two constants
\beqar
\gamma^+ &=& \frac{1}{4} \left(2\hbar\om^+ - g\mu_B B\right)\\
\zeta'(\omega) &=& - \lambda\sq{{m\hbar\om\over2}}
\left(1+{\om_c\over\om}\right)
\eeqar
we write the Hamiltonian $H_d$ \eqref{hamd}
as
\beq\lb{n0}
H_d=\hbar\omega^{+}
M_d-\gamma^+\sigma_z+\zeta'(\omega)\left(a_d^\dag\sigma^{-}+a_d\sigma^{+}\right).
\eeq
Doing the same job to find the required solutions
\begin{eqnarray}
\sigma^{+}(t) &=&{e^{i\omega^+t}\ov
\al_d^--\al_d^+}\left\{\left(\zeta'(\omega)
a_d^{\dag}(0)-\al_d^+\si^+(0)\right)e^{i{\al_d^{+}t/\hbar}}+\left(-\zeta'(\omega)
a_d^{\dag}(0)+\al_d^+\si^+(0)\right)e^{i{\al_d^{-}t/\hbar}}\right\}\lb{3}\\
a_d^{\dag}(t) &=& {e^{i\omega^+t}\ov
\al_d^--\al_d^+}\left\{\left(-\zeta'(\omega)
\si^{+}(0)+\al_d^-a_d^{\dag}(0)\right)e^{i{\al_d^{+}t/\hbar}}+\left(\zeta'(\omega)
\si^{+}(0)-\al_d^+a_d^{\dag}(0)\right)e^{i{\al_d^{-}t/\hbar}}\right\}\lb{dso2}
\end{eqnarray}
where different quantities are given by
\beqar
\al_d^{\pm} &=& C_d\pm\sqrt{C_d^{2}-\zeta'^2(\omega)}\\
C_d &=&-\gamma^+\sigma_z+\zeta'(\omega) \left(a_g^\dag\sigma^{-}+a_g\sigma^{+}\right).
\eeqar
These solutions are showing how to obtain the second copy of the Jaynes-Cummings model
from our findings. In summary, we conclude that our Hamiltonian  is equivalent to two copies
of   the Jaynes-Cummings model but oscillating with different frequencies.

We close this part by noting that,
 we can easily obtain the dynamic
of the complex position from the above solutions. Indeed, using \eqref{xyrep} to write
\beq
z(t)= l_0\left(a_d^\dag(t) + a_g(t) \right)
\eeq
and therefore summing up the adjoint time evolution operator of \eqref{33} and \eqref{dso2} to end
up with the dynamics of $z(t)$. The established link shows clearly that our system is sharing some common
features with quantum optics. Thus, one can use the present system to handle different issues related
to Jaynes-Cummings model and vice vera.

\subsection{Limiting case}

The above results show
the analogy to Jaynes-Cummings model
and therefore allow us to establish a relation with the already published work~\cite{ackerhalt}.
Now, we study the case where the confining potential is absent,
which naturally  should lead to  the Schliemann
results~\cite{schliemann} for Jaynes-Cummings model.
These will be derived in simple way from our findings to show
clearly that our work is general and can be extended to deal with different issues.

To recover the dynamics obtained in~\cite{schliemann} for Jaynes-Cummings model,  we start our job by fixing the frequency
as $\omega_0=0$ in the previous results. This requirement
leads to the constraints
\beq
\om^-=0, \qquad l_1=2, \qquad l_2=0
\eeq
and therefore according to the dynamical equation \eqref{444}
or (\ref{33}) we end up with
an operator $a^\dag_g$ time independent
\beq
a_g^{\dag}(t)=a_g^{\dag}(0)=a_g^{\dag}.
\eeq
However, (\ref{dso2}) gives $a_d^{\dag}(t)$ time dependent
\begin{equation}
a_d^{\dag}(t)={e^{-i\omega_ct}\ov
r_d^+-r_d^-}\left\{\left(\zeta'(\omega_c) \si^{+}(0)-r_d^-a_d^{\dag}(0)\right)e^{-i{r_d^{+}\ov\hbar}t}+\left(-\zeta'(\omega_c)
\si^{+}(0)+r_d^+a_d^{\dag}(0)\right)e^{-i{r_d^{-}\ov\hbar}t}\right\}
\end{equation}
where the roots are
\beq
r_d^{\pm}=C_d\pm{\left(C_d^{2}+\zeta'^2(\omega_c)\right)^{{1\ov2}}}
\eeq
and all involved functions are now in terms of the cyclotron frequency $\om_c$ instead of $\om$. Returning back
to the definition of different operators to obtain
the time evolution of the position operators in the Heisenberg picture. This simply is 
\beq
x(t)+iy(t)=l_0\left(a_d^{\dag}(t)+a_g\right).
\eeq
Since the operators  $a_g$ is time independent, then at $t=0$ we have
\beq
x(0)+iy(0)=l_0a_g
\eeq
and then after replacing, we find
\beq
x(t)+iy(t)=x(0)+iy(0)+l_0a_d^{\dag}(t).
\eeq
From \eqref{pieq} we can express $a_d^{\dag}$ in terms of the
conjugate momentum as
\beq
a_d^{\dag}={l_0\ov2i\hbar}\left(\pi_x+i\pi_y\right)
\eeq
which leads to the final form of the complex position
\begin{eqnarray}\lb{01}
x(t)+iy(t) &=& x(0)+iy(0)+i{e^{-i(\omega_c+{r_d^+\ov\hbar})t}\ov r_d^+-r_d^-}\left({r_d^-\ov\om_c}{\pi_x+i\pi_y\ov m}-i2\lambda\hbar\sigma^+\right)\nonumber\\
&&-i{e^{-i(\omega_c+{r_d^-\ov\hbar})t}\ov r_d^+-r_d^-}\left({r_d^+\ov\om_c}{\pi_x+i\pi_y\ov m}-i2\lambda\hbar\sigma^+\right)
\end{eqnarray}
where
the operators valued $r_d^{\pm}$ are given by
\beq
r_d^{\pm}=C_d\pm{\left(C_d^{2}+2\lambda^2m\hbar\omega_c\right)^{{1\ov2}}}.
\eeq
This nothing but the result obtained by Schliemann~\cite{schliemann} in dealing with the same system
without confinement. Thus, it really shows
that our findings are important as well as general in sense that we can derive other results.

\section{Conclusion}

We have investigated the basic features of confined two-dimensional
system with Rashba spin-orbit interaction in the
presence of an external magnetic field $B$. This latter allowed us
to end up with a confining potential along $x$ and $y$-directions that
has been used to deal with different issues. In particular, it has
been served to split the corresponding Hamiltonian into two parts.
This decomposition was useful in sense that different spectrum
are obtained and lead to the total solutions of
the energy spectrum. We have shown that those
obtained by Schliemann \cite{schliemann} can be derived
in the simple way from our solutions.

Using different operators involved in the Hamiltonian, we have
realized two dynamical symmetries $su(2)$ and $su(1,1)$. These
together with the strength of magnetic field $B$ allowed us to discuss
the {filling} of the shells with electrons. For strong $B$, we have
concluded that our system behaves like a harmonic oscillators in 2D
with an infinite degeneracy of the Landau levels. However, for {weak}
$B$ our system becomes invariant under the algebra $su(2)$ and then
each Landau level has an finite degeneracy.

To make contact with quantum optics, we have elaborated
a method based on building two Hamiltonian's from the original one.
Indeed, by splitting this later into tow parts we have shown that
it is possible to recover the Jaynes-Cummings model, which is describing
a system with two states. This has been done by using the Heisenberg
dynamics to find the dynamics of the raising Pauli operator $\si^+$ and
creation operators $(a^\da_d,a^\da_g)$. After solving different equations,
we have ended up with the exact operator solutions for the basic
Jaynes-Cummings variables those have been obtained in \cite{ackerhalt}.
To show the validity of our results, we have derived those
obtained by Schliemann \cite{schliemann} as particular cases.

The present work can be extended to deal with different issues. For instance,
we can use the  route used by Schliemann \cite{schliemann} to explicitly study
the full quantum dynamics. This is based on expanding the initial state of the system
in terms of its eigenstates. Another alternative is to use the obtained results
to study different issues related to graphene and spin Hall effect.

\section*{Acknowledgments}

The generous support provided by the Saudi Center for Theoretical Physics (SCTP) is highly appreciated by AJ.
He also acknowledges the support provided by King Faisal University. {We thank the referees for
their comments.}

\end{document}